\documentclass{sig-alternate}
\usepackage{graphicx}
\usepackage{amsmath}
\usepackage{bbm}
\usepackage{amssymb}
\usepackage{epstopdf}
\usepackage{helvet}
\usepackage{enumitem}

\begin{document}
\title{Open Packet Processor: a programmable architecture for wire speed platform-independent stateful in-network processing}

\author{
Giuseppe Bianchi, Marco Bonola, Salvatore Pontarelli,\\Davide Sanvito, Antonio Capone, Carmelo Cascone\\
\affaddr{CNIT / University of Rome Tor Vergata / Politecnico di Milano}
}

\maketitle

\begin{abstract}
This paper aims at contributing to the ongoing debate on how to bring programmability of stateful packet processing tasks inside the network switches, while retaining platform independency. Our proposed approach, named ``Open Packet Processor'' (OPP), shows the viability (via an hardware prototype relying on commodity HW technologies and operating in a strictly bounded number of clock cycles) of eXtended Finite State Machines (XFSM) as low-level data plane programming abstraction. With the help of examples, including a token bucket and a C4.5  traffic classifier based on a binary tree, we show the ability of OPP  to support stateful operation and flow-level feature tracking. Platform independence is accomplished by decoupling the implementation of hardware primitives (registries, conditions, update instructions, forwarding actions, matching facilities) from their usage by an application formally described via an abstract XFSM. We finally discuss limitations and extensions.
\end{abstract}


\section{Introduction}
\label{s:intro}
To face the emerging needs for service flexibility, network efficiency, traffic diversification, and network security and reliability, today's network nodes are called to a more flexible and richer packet processing. The original Internet nodes, historically limited to switches and routers providing ``just'' the plain forwarding services, have been massively complemented with a variety of heterogeneous middlebox-type functions \cite{Wan11,She12,Qaz13,Gem14} such as network address translation, tunneling, load balancing, monitoring, intrusion detection, and so on. The diversification of network equipment and technologies has definitely provided an increased availability of network functionalities, but at the cost of a significant extra complexity in the control and management of large scale multi-vendor networks.

Software-defined Networking (SDN) emerged as an attempt to address such problem. Coined in 2009 \cite{Gre09} as a direct follow-up of the OpenFlow proposal \cite{OF08}, SDN has broadly evolved since then \cite{frz14} and does not in principle restricts to OpenFlow (a {\em ``minor piece in the SDN architecture''}, according to the OpenFlow inventors themselves \cite{She11}) as device-level abstraction. Nevertheless, most of the high level network programming abstractions proposed in the last half a dozen years \cite{Gud08,Nay09,Fos11,Voe12,Mon13,Nel14,Kim15} still rely on OpenFlow as southbound (using RFC 7426's terminology) programming interface. Indeed, OpenFlow was designed with the desire for rapid adoption, opposed to first principles \cite{frz14}; i.e., as a pragmatic attempt to address the dichotomy between i) flexibility and ability to support a broad range of innovation, and ii) compatibility with commodity hardware and vendors' need for closed platforms \cite{OF08}. 

The aftermath is that most of the above mentioned network programming frameworks circumvent OpenFlow's limitations by promoting a ``two-tiered'' \cite{Ara15} programming model: {\em any} stateful processing intelligence of the network applications is delegated to the network controller, whereas OpenFlow switches limit to install and enforce stateless packet forwarding rules delivered by the remote controller. Centralization of the network applications' intelligence may not be a problem (and actually turns out to be an advantage) whenever changes in the forwarding states do not have strict real time requirements, and depend upon global network states. But for applications which rely only on local flow/port states, the latency toll imposed by the reliance on an external controller rules away the possibility to enforce software-implemented control plane tasks at wire speed, i.e. while remaining on the fast path\footnote{
A 64 bytes packet takes about 5 ns on a 100 gbps speed, roughly the time needed for a signal to reach a control entity placed one meter away. And the execution of an albeit simple software-implemented control task may take way more time than this. Thus, even the physical, capillary, distribution of control agents (as proxies of the remote SDN controller for low latency tasks) on each network device would hardly meet fast path requirements. }. 

One might argue that we {\em do not even need} such ultra-fast processing and packet-by-packet manipulation and control capabilities. However, not only the large real-world deployment of proprietary hardware network appliances (e.g. for traffic classification, control/balancing, monitoring, etc), but also the evolution of the OpenFlow specification itself shows that this may not be the case. As a matter of fact, since the creation of the Open Networking Foundation (ONF) in 2011, and up to the latest (version 1.5) specification, we have witnessed an hectic evolution of the OpenFlow standard, with several OpenFlow extensions devised to fix punctual shortcomings and accommodate specific needs, by incorporating {\em extremely specific} stateful primitives (such as meters for rate control, group ports for fast failover support or dynamic selection of one among many action buckets at each time - e.g. for load balancing -, synchronized tables for supporting learning-type functionalities, etc). 

Indeed, in the last couple of years, a new research trend has started to challenge improved programmability of the data plane, beyond the elementary ``match/action'' abstraction provided by OpenFlow, and (even more recently) initial work on higher level network programming frameworks devised to exploit such newer and more capable lower-level primitives are starting to emerge \cite{Sha15,Ara15}. Proposals such as POF \cite{Son13,Son15}, although not yet targeting stateful flow processing, do significantly improve header matching flexibility and programmability, freeing it from any specific structure of the packet header. Programmability of the packet scheduler inside the switch has been recently addressed in \cite{Siv15}. Works such as OpenState \cite{ccr14,Pon15} and FAST \cite{Mos14} explicitly add support for per-flow state handling inside OpenFlow switches, although the abstractions therein defined are still simplistic and severely limit the type of applications that can be deployed (for instance, OpenState supports only a special type of Finite State Machines, namely Mealy Machines, which do not provide the programmer with the possibility to declare and use own memory or registries). The P4 programming language \cite{Bos14,Jos15}  leverages more advanced hardware technology, namely dedicated processing architectures \cite{flexpipe} or Reconfigurable Match Tables \cite{Bos13} as an extension of TCAMs (Ternary Content Addressable Memories) to permit a significantly improved programmability in the packet processing pipeline. In its latest 1.0.2 language specification \cite{P4spec}, P4 has made a further crucial step in improving stateful processing, by introducing registers defined as {\em ``stateful memories} [which] {\em can be used in a more general way to keep state''}. However, the P4 language does not specify how registries should be scalably supported and managed by the underlying HW.

\vspace{-3pt}
\subsubsection*{Contribution}

This work is an attempt to revisit fast-path programmability, by (i) proposing a programming abstraction which retains the platform independent features of the original ``match/action'' OpenFlow abstraction, and by (ii) showing how our abstraction can be directly ``executed'' over an HW architecture (whose feasibility is concretely shown via an HW FPGA prototype). In analogy with the OpenFlow's ``match/action'' abstraction, which exposes a network node's TCAM to third party programmability, also our abstraction directly refers to the HW interface, and as such it can be directly exposed to the programmer as a machine-level ``configuration'' interface, hence without any intermediary compilation or adaptation to the target (i.e., unlike the case of P4). 

In conceiving our abstraction, we have been largely inspired by \cite{ccr14}, where eXtended Finite State Machines \cite{Che93} (therein referred to as ``full'' XFSM) were conjectured as a possible forward-looking abstraction. Our key difference with respect to \cite{ccr14} is that we not limit to postulate that such ``full'' XFSMs may ultimately be a suitable abstraction, but we concretely show their viability and their ``executability'' over an HW architecture leveraging commodity HW (standard TCAMs, hash tables, ALUs, and somewhat trivial additional circuitry), and with a strictly bounded number of clock ticks.

\begin{figure}[t]
	\includegraphics[width=0.5\textwidth]{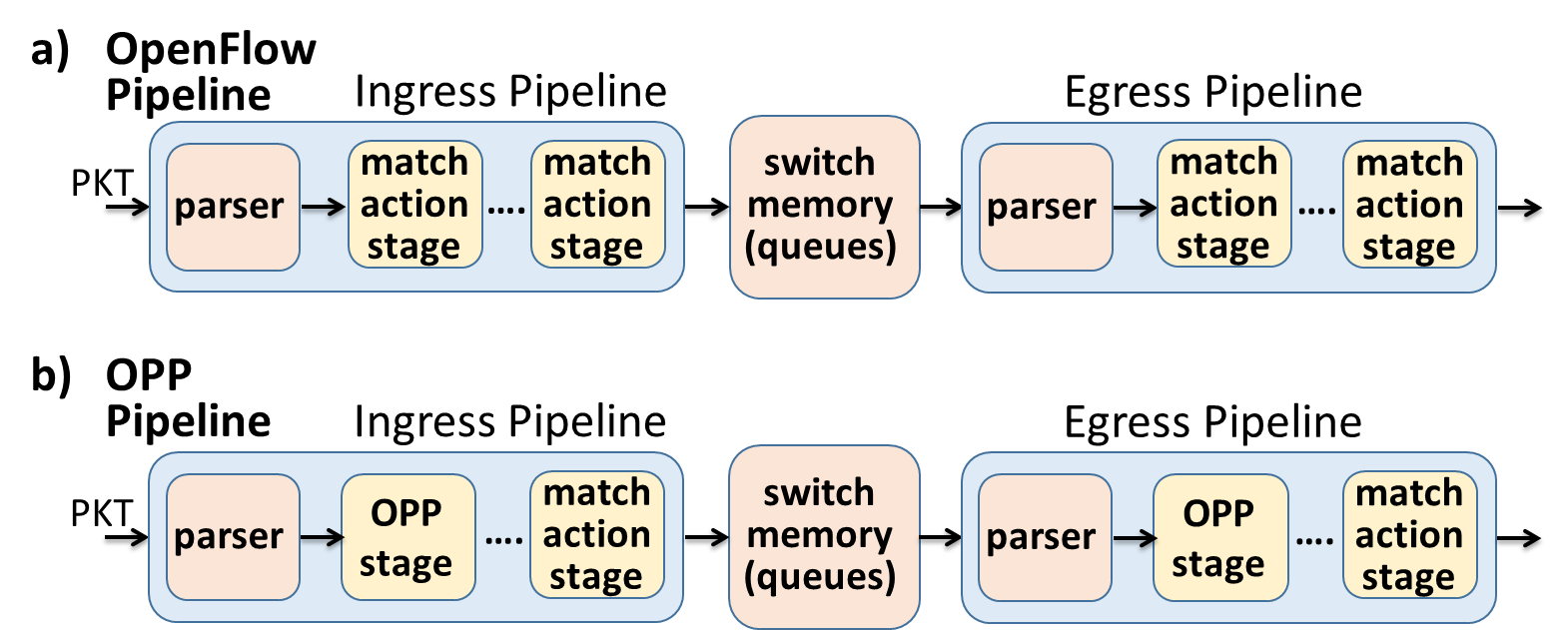}
	  \vspace{-2em}
	\caption{a) A typical OpenFlow pipeline architecture. b) the OPP enabled pipeline. OPP ``stages'' can be pipelined with other OPP stages or ordinary OpenFlow Match/Action stages.}
	  \vspace{-1em}
	\label{f:pipe}
\end{figure}

A limitation in this paper is our focus on a ``single'' packet processing stage, opposed to a more general packet processing pipeline comprising multiple match/action tables. In essence, our work shows the viability of an XFSM-based abstraction as a (significant) generalization of the original single-table OpenFlow's match/action. While multiple pipelined instances of our atomic Open Packet Processor stages are clearly possible, exactly as multiple match/action tables can be pipelined since OpenFlow version 1.1 (see figure \ref{f:pipe}), our present work does not yet take advantage of HW pipeline optimizations such as Reconfigurable Match Tables \cite{Bos13}.

Finally, and similarly to the OpenFlow's original design philosophy, even if our proposed architecture is pragmatically  limited by the specific set of primitives implemented by the HW (supported packet processing and forwarding actions, matching facilities, arithmetic and logic operations on registry values, etc), it nevertheless remains extensible (by adding new actions or instructions) and largely expressive in terms of how the programmer shall use and combine such primitives within a desired stateful operation. As it will be hopefully apparent later on, a ``full'' XFSM permits to formally describe a wide variety of programmable packet processing and control tasks, which our architecture permits to directly convey and deploy {\em inside} the switch. And even if probably not of any practical interest, the fact that not only the OpenFlow legacy statistics, but also further tailored stateful extensions today integrated in the OpenFlow standard (hence hardcoded in the switch) might be {\em externally} programmed using an apparently viable platform agnostic abstraction merits further considerations (see discussion in section \ref{s:disc}).


\section{Concept}
\label{s:concept}

As anticipated in the previous section, our work focuses on the design of a single Open Packet Processor (OPP) stage, as a significant generalization of the traditional OpenFlow's Match/Action abstraction. More specifically, our goal is to provide a packet processing stage which holds the following properties.

\vspace{3pt} \noindent {\bf Ability to process packets directly on the fast path}, i.e., {\em while} the packet is traveling in the pipeline (nanoseconds time scale). The requirement of performing packet processing tasks in a {\em deterministic} and {\em small} (bounded) number of HW clock cycles hardly copes with the possibility to employ a standard CPU (and the relevant programming language), and requires us to implement a domain-specific (traffic/network) computing architecture from scratch. 
 
\vspace{3pt} \noindent {\bf Efficient storage and management of per-flow stateful information}. Other than parsing packet header information and exposing such fields to a match/action stage (or a pipeline of match action stages\cite{Bos13,flexpipe}), we also aim at permitting the programmer to further use the {\em past} flow history for defining a desired per-packet processing behaviour. As shown in section \ref{ss:arch}, this can easily accomplished by {\em pre-pending} a dedicated storage table (concretely, an hash table) that permits to retrieve, in O(1) time, stateful flow information. We name this structure as {\em Flow Context Table}, as, in somewhat analogy with context switching in ordinary operating systems, it permits to retrieve stateful information associated to the flow to which an arriving packet belongs, and store back an updated context the end of the packet processing pipeline. Such (flow) context switching will operate at wire speed, on a packet-by-packet basis. 

\vspace{3pt} \noindent {\bf Ability to specify and compute a wide (and programmable) class of stateful information}, thus including counters, running averages, and in most generality stateful features useful in traffic control applications. It readily follows that the packet processing pipeline, which in standard OpenFlow is limited to match/action primitives, must be enriched with means to describe and (on the fly) enforce conditions on stateful quantities (e.g. the flow rate is above a threshold, or the time elapsed since the last seen packet is greater than the average inter-arrival time), as well as provide arithmetic/logic operations so as to update such stateful features in a bounded number of clock cycles (ideally one). 

\vspace{3pt} \noindent {\bf Platform independence}. A key pragmatic insight in the original OpenFlow abstraction was the decision of restricting the OpenFlow switch programmer's ability to just {\em select} actions among a finite set of supported ones (opposed to permitting the programmer to develop own custom actions), and associate a desired action set (bundle) to a specific packet header match. We conceptually follow a similar approach, but we cast it into a more elaborate eXtended Finite State Machine (XFSM) model. As described in section \ref{ss:xfsm}, an XFSM abstraction permits us to formalize complex behavioral models, involving custom per-flow states, custom per-flow registers, conditions, state transitions, and arithmetic and logic operations. Still, an XFSM model does not require us to know {\em how} such primitives are concretely implemented in the hardware platform, but ``just'' permits us to combine them together so as to formalize a desired behaviour. Hence, it can be {\em ported} across platforms which support a same set of primitives.

\subsection{XFSM abstraction}
\label{ss:xfsm}
The OpenFlow's ``Match-action'' abstraction has been widely extended throughout the various standardization steps, with the extension of the match fields (including the possibility to perform matches on meta-data), with new actions (and instructions), and with the ability to associate a set of actions to a given match. Nevertheless, the basic abstraction conceptually remains the same. It is instructive to formally re-interpret the (basic) OpenFlow match/action abstraction as a ``map'' $T : I \rightarrow O$, where $I=\{i_1, \ldots, i_M\}$ is a finite set of {\em Input Symbols}, namely all the possible matches which are technically supported by an OpenFlow specification (being irrelevant, at least for this discussion, to know how such Input Symbols' set $I$ is established, and that each input symbol is a Cartesian combination of all possible header field matches), and $O=\{o_1, \ldots, o_K\}$ is a finite set of {\em Output Symbols}, i.e. all the possible actions supported by an OpenFlow switch. The obvious limit of this abstraction is that the match/action mapping is statically configured, and can change only upon controller's intervention (e.g. via flow-mod OpenFlow commands). Finally, note that the ``engine'' which performs the actual mapping $T : I \rightarrow O$ is a standard TCAM.

As observed in \cite{ccr14}, an OpenFlow switch can be trivially extended to support a more general abstraction which takes the form of a {\em Mealy Machine}, i.e. a Finite State Machine with output,
and which permits to formally model {\em dynamic} forwarding behaviors, i.e. permit to change in time the specific action(s) associated to a same match. It suffices to add a further finite set $S=\{s_1, s_2, …, s_N\}$ of {\em programmer-specific states}, and use the TCAM to perform the mapping $T : S \times I \rightarrow S \times O$. While remaining feasible on ordinary OpenFlow hardware, such Mealy Machine abstraction brings about two key differences with respect to the original OpenFlow abstraction. First, the (output) action associated to a very same (input) match may now differ depending on an (input) {\em state} $s_i \in S$, i.e., the state in which the flow is found when a packet is being processed. Second, the Mealy Machine permits to specify in which, possibly different, (output) state $s_o \in S$ the flow shall enter once the packet will be processed. While quite interesting, this generalization appears still insufficient to permit the programmer to implement meaningful applications, as it lacks the ability to run-time compute and exploit in the forwarding decisions per-flow features commonly used in traffic control algorithms. 

\begin{table}[t]
\centering 
{\footnotesize
\begin{tabular}{|c|p{2.9cm}|p{4.2cm}|} \hline
\multicolumn{2}{|c|} {\bf XFSM formal notation}  & 
			{\bf Meaning} \\ \hline
I & input symbols & all possible matches on packet header fields  \\ \hline
O & output symbols & OpenFlow-type actions \\ \hline
S & custom states & application specific states, defined by programmer \\ \hline
D & n-dimensional linear space $ D_1 \times \cdots \times D_n $ & 
			all possible settings of $n$ memory registers; 
			include both custom per-flow and global switch registers (see text) \\ \hline
F & set of enabling functions $f_i: D \rightarrow \{0,1\}$ & 
			Conditions (boolean predicates) on registers \\ \hline
U & set of update functions $u_i: D \rightarrow D$ &
			Applicable operations for updating registers' content \\ \hline
  T & transition relation $T: S \times F \times I \rightarrow 
			S \times U \times O $ &
			Target state, actions and register update  
			commands associated to each transition \\ \hline
\end{tabular}}
\caption{eXtended Finite State Machine model}
      	  \vspace{-1.2em}
\label{t:xfsm}
\end{table}

The goal of this paper is to show that a switch architecture can be further easily extended (section \ref{ss:arch}) to support an even more general Finite State Machine model, namely the eXtended Finite State Machine (XFSM) model introduced in \cite{Che93}. As summarized in table \ref{t:xfsm}, this model is formally specified by means of a 7-tuple $M=(I,O,S,D,F,U,T)$. Input symbols $I$ (OpenFlow-type matches) and Output Symbols $O$ (actions) are the same as in OpenFlow. Per-application states $S$ are inherited from the Mealy Machine abstraction \cite{ccr14}, and permit the programmer to freely specify the possible states in which a flow can be, in relation to her desired custom application (technically, a state label is handled as a bit string). For instance, in an heavy hitter detection application, a programmer can specify states such as \texttt{NORMAL}, \texttt{MILD}, or \texttt{HEAVY}, whereas in a load balancing application, the state can be the actual switch output port number (or the destination IP address) an already seen flow has been pinned to, or \texttt{DEFAULT} for newly arriving flows or flows that can be rerouted. With respect to a Mealy Machine, the key advantage of the XFSM model resides in the additional  programming flexibility in three fundamental aspects. 

\vspace{3pt} \noindent {\bf (1) Custom (per-flow) registers and global (switch-level) parameters.} The XFSM model permits the programmer to explicitly define her own registers, by providing an array of per-flow variables whose content (time stamps, counters, average values, last TCP/ACK sequence number seen, etc) shall be decided by the programmer herself. Additionally, it is useful to expose to the programmer (as further registers) also switch-level states (such as the switch queues' status) or ``global'' shared variables which all flows can access. Albeit practically very important, a detailed distinction into different register types is not foundational in terms of abstraction, and therefore all registers that the programmer can access (and eventually update) are summarized in the XFSM model presented in Table \ref{t:xfsm} via the {\em array D} of {\em memory registers}.

\vspace{3pt} \noindent {\bf (2) Custom conditions on registers and switch parameters.} The sheer majority of traffic control applications rely on {\em comparisons}, which permit to determine whether a counter exceeded some threshold, or whether some amount of time has elapsed since the last seen packet of a flow (or the first packet of the flow, i.e., the flow duration). The {\em enabling functions} $f_i:D\rightarrow\{0,1\}$ serve exactly for this purpose, by implementing a set of (programmable) boolean comparators, namely conditions whose input can be decided by the programmer, and whose output is 1 or 0, depending on whether the condition is true or false. In turns, the outcome of such comparisons can be exploited in the transition relation, i.e. a state transition can be triggered only if a programmer-specific condition is satisfied. 

\vspace{3pt} \noindent {\bf (3) Register's updates.} Along with the state transition, the XFSM models also permits the programmer to update the content of the deployed registers. As we will show later on, registers'  updates require the HW to implement a set of {\em update functions} $u_i:D\rightarrow D$, namely arithmetic and logic primitives which must be provided in the HW pipeline, and whose input and output data shall be configured by the programmer.

\vspace{3pt} Finally, we stress that the actual computational step in an XFSM is the transition relation $T : S \times F \times I \rightarrow S \times U \times O$, which is nothing else than a ``map'' (albeit with more complex inputs and outputs than the basic OpenFlow map), and hence is naturally implemented by the switch TCAM, as shown in the next section \ref{ss:arch}.

\begin{figure*}[t]
	\centering
	\includegraphics[width=1\textwidth, height=5cm]{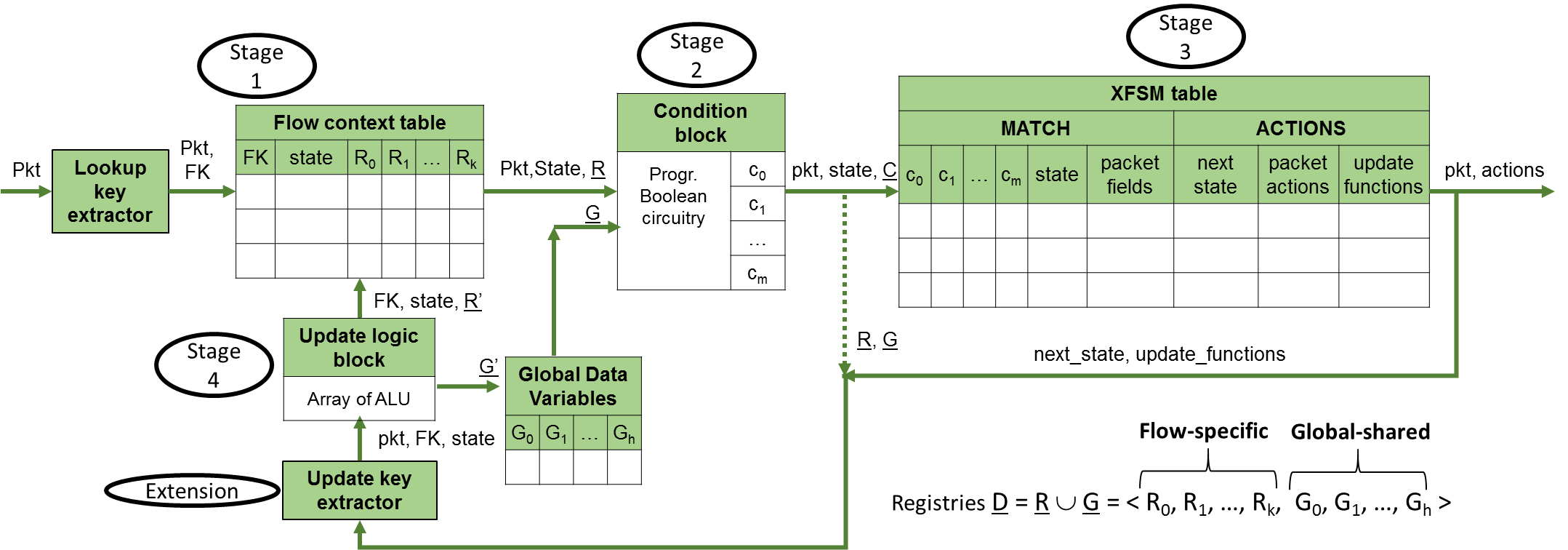} 
		  \vspace{-2em}
	\caption{OPP architecture}
		  \vspace{-1em}
	\label{f:hl_arch}
\end{figure*}

\subsection{OPP architecture}
\label{ss:arch}

To our view, what makes the previously described XFSM abstraction compelling is the fact that it can be {\em directly executed on the switch's fast path} using off the shelf HW, as we will prove in section \ref{s:hardware} with a concrete HW prototype. As discussed in the next section, practical restrictions of course emerge in terms of memory deployed for the registers, as well as capability of the ALUs used for register updates, but such restrictions are mostly related to an actual {\em  implementation}, rather than to the design which remains at least in principle very general and flexible. A sketch of the proposed Open Packet Processor architecture is illustrated in Figure \ref{f:hl_arch}. The packet processing workflow is best explained by means of the following {\em stages}.

\vspace{3pt} \noindent {\bf Stage 1: flow context lookup}. Once a packet enters an OPP processing block, the first task is to extract, from the packet, a {\em Flow Identification Key} (FK), which identifies the entity to which a state may be assigned. The flow is identified by an unique key composed of a subset of the information stored in the packet header. The desired FK is configured by the programmer  (an IP address, a source/destination MAC pair, a 5-tuple flow identifier, etc) and depends on the specific application. The FK is used as index to lookup a {\em Flow Context Table}, which stores the {\em flow context}, expressed in terms of (i) the state label $s_i$ currently associated to the flow, and (ii) an array $\vec{R} = \{R_0, R_1, ..., R_k\}$ of (up to) $k+1$ registers defined by the programmer. The retrieved flow context is then appended as metadata and the packet is forwarded to the next stage.

\vspace{3pt} \noindent {\bf Stage 2: conditions' evaluation}. Goal of the {\em Condition Block} illustrated in Figure \ref{f:hl_arch} (and implemented using ordinary boolean circuitry, see section \ref{s:hardware}) is to compute programmer-specific conditions, which can take as input either the per flow register values (the array $\vec{R}$), as well as global registers delivered to this block as an array $\vec{G} = \{G_0, G_1, ..., G_h\}$ of (up to) $h+1$ global variables and/or global switch states. Formally, this block is therefore in charge to implement the {\em enabling functions} specified by the XFSM abstraction. In practice, it is trivial to extend the assessment of conditions also to packet header fields (for instance, port number greater than a given global variable or custom per-flow register). The output of this block is a boolean vector $\vec{C} = \{c_0, c_1, ..., c_m\}$ which summarizes whether  the $i$-th condition is true ($c_i=1$) or false ($c_i=0$). 

\vspace{3pt} \noindent {\bf Stage 3: XFSM execution step}. Since boolean conditions have been transformed into 0/1 bits, they can be provided as input to the TCAM, along with the state label and the necessary packet header fields, to perform a wildcard matching (different conditions may apply in different states, i.e. a bit representing a condition can be set to ``don't care'' for some specific states). Each TCAM row models one transition in the XFSM, and returns a 3-tuple: (i) the next state in which the flow shall be set (which could coincide with the input state in the case of no state transition, i.e., a self-transition in the XFSM), (ii) the actions associated the transition (usual OpenFlow-type forwarding actions, such as \texttt{drop}, \texttt{push\_label}, \texttt{set\_tos} etc...), and (iii) the information needed to update the registers as described below. 

\vspace{3pt} \noindent {\bf Stage 4: register updates}. Most applications require arithmetic processing when updating a stateful variable. Operations can be as simple as integer sums (to update counters or byte statistics) or can require tailored floating point processing (averages, exponential decays, etc). The role of the {\em Update logic block} component highlighted in Figure \ref{f:hl_arch} is to implement an array of Arithmetic and Logic Units (ALUs) which support a selected set of computation primitives which permit the programmer to update (re-compute) the value of the registers, using as input the information available at this stage (previous values of the registers, information extracted from the packet, etc). Section \ref{ss:alu} will describe the specific instruction set implemented in our HW prototype, where (with no pretence of completeness, nor willingness to impose our own set of operations) we implement a set of operations which appear to be either useful to the specific network programmer's needs, as well as computationally effective in terms of implementation (ideally, executable in a single clock tick). It is worth to mention that the problem of extending the set of supported ALU instructions is merely a technical one, and does not affect the OPP architecture.  

\vspace{3pt} \noindent {\bf Extension: Cross-flow context handling}. As noted in \cite{ccr14}, there are many useful stateful control tasks, in which states for a given flow are updated by events occurring on {\em different} flows. A simple but prominent example is MAC learning: packets are forwarded using the {\em destination} MAC address, but the forwarding database is updated using the {\em source} MAC address. Thus, it may be useful to further generalize the XFSM abstraction as suggested in \cite{ccr14}, i.e. by permitting the programmer to use a Flow Key during lookup (e.g. read information associated to a MAC destination address) and employ a possibly different Flow Key (e.g. associated to the MAC source) for updating a state or a register.

\subsection{Programming the OPP}

Is is useful to conclude this section with at least a sketch of which types of applications (and programs) may be deployed. 

A first trivial example of dynamic forwarding actions is that of a simple mechanism which distinguishes {\bf short-lived flows from long-lived flows} by considering ``long'' any flow that has transmitted at least N packets, and applies different DSCP tags. The OPP programmer would simply need to define two states (\texttt{DEFAULT} also associated to every new flow, and \texttt{LONG}), one per-flow register $R_0$ (a packet counter), one global register $G_0$ (storing the constant threshold $N$), a condition $R_0>G_0$ applicable when in state \texttt{DEFAULT}, and an update function ${\rm ADD}(R_0,1) \rightarrow R_0$. Note that we {\em did not} assume any pre-implemented counters or meters in the switch, but the counter and the relevant threshold check has been programmed using the OPP abstraction.

The usage of packet inter-arrival times and timers is exemplified by a {\bf dynamic intra-flow load balancing} application, which can reroute a flow {\em while} it is in progress. As suggested in \cite{Kan07}, rerouting should {\em not} occur during packet bursts, to avoid out of ordering and relevant performance impairments. Support in OPP just requires, for each packet being transmitted, to update a per-flow register $R$ with the quantity $t+\Delta$, being $t$ the actual packet timestamp and $\Delta$ a suitable threshold. When the next packet arrives (time $t_1$), we check the condition $t_1>R$. If this is false, we route the packet to the assigned path (indicated by the state label); conversely we route it to an alternative path, and we change the state accordingly. Again, note that we have not assumed any pre-implemented support from the switch (e.g. timeouts or soft states), besides the ability to provide time information (e.g., via a global register, or timestamps as packet metadata).

Finally, the integration in the ALU design of monitoring-specific update instructions (averages, variances, smoothing filters, see section \ref{ss:alu}), several features frequently used in traffic control and classification applications can be computed on the fly during the pipeline. We defer relevant examples to section \ref{s:usecases}.


\section{Hardware feasibility}
\label{s:hardware}

Despite the current trend in softwarization of network functions and the widespread deployment of software switches, we believe that the viability of switch-level programming abstractions which challenge OpenFlow limitations, hence including this work, {\em must} still be proven in terms of hardware feasibility and ability to run in a strictly bounded number of clock cycles. 

\begin{figure}[t]
\centering
   \includegraphics[width=0.52\textwidth]{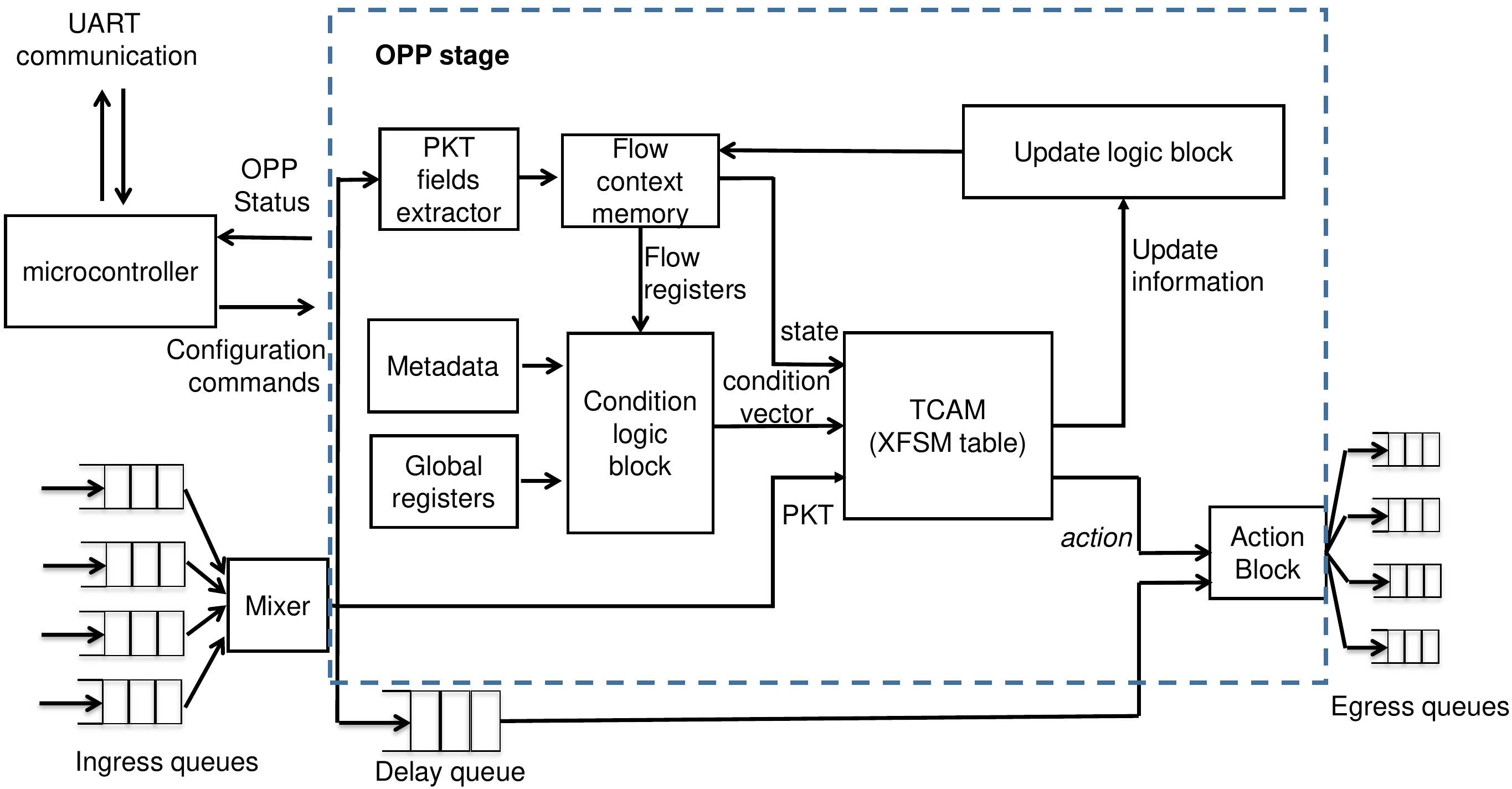}
   	  \vspace{-2em}
\caption{Scheme of an OPP stage}
	  \vspace{-1.3em}
\label{F:HW_core}
\end{figure}

Figure \ref{F:HW_core} provides a block-level overview of a candidate hardware implementation of a single OPP stage, which we prove feasible with an FPGA prototype. Pipelining of an OPP stage with other OPP stages or ordinary match/action tables does not affect the single stage design (although it does not permit us to benefit from hardware extensions and TCAM optimizations such as those introduced in \cite{Bos13}, which we leave to future work). Figure \ref{F:HW_core} also illustrates the necessary auxiliary blocks devised to handle packet input capture and output delivery, in the assumption of a $4 \times 4$ port switch.

\vspace{3pt} \noindent {\bf Packet reception and header field extraction}.
Packets received on the input queues are collected and serialized by a mixer block, so that the OPP block receives one packet per clock cycle. Such packet is then processed by a {\em Packet Fields Extractor}, configured to provide, together  with the header fields (8 in our prototype), the blocks required in the next processing stages - specifically: i) the Flow Key used to query the Flow Context Table, ii) the header fields used by the Condition block, iii) the header fields used by the Update Logic Block, and iv) the (eventually different) Flow Key used for updating the Flow Context. The Packet Fields Extractor is easily implemented in HW as a parallel array of elementary Shift and Mask (SaM) blocks where each SaM block selects the beginning of the targeted header field (the shift function), and performs a bit-wise mask operation. This operation closely resembles that proposed in POF \cite{Son13}: we also use offsets, but instead of lengths we use bit masks.  

\vspace{3pt} \noindent {\bf Flow Context Table}. This data structure is in charge to store both state as well as registries associated to Flow Keys. It consists of an hash table (we implemented a  {\em d-left hash table} with $d=4$) to handle exact matches, plus a TCAM to handle wildcard matches. Unlike the hash table, which must be arguably large to store per-flow states, a very small TCAM can be deployed, as it is required to handle the very few special cases where wildcard matches are needed (mainly default states, where the TCAM priority permits to differentiate default states for different protocols or packet formats). Our implementation uses 128 bit Flow Keys, and returns a 146 bit value which is sufficient to support a 16 bit state label, four 32 bit per-flow registries, and two auxiliary bits per entry used by the microcontroller for housekeeping (see below). 

\begin{figure}[t]
\centering
   \includegraphics[width=0.4\textwidth]{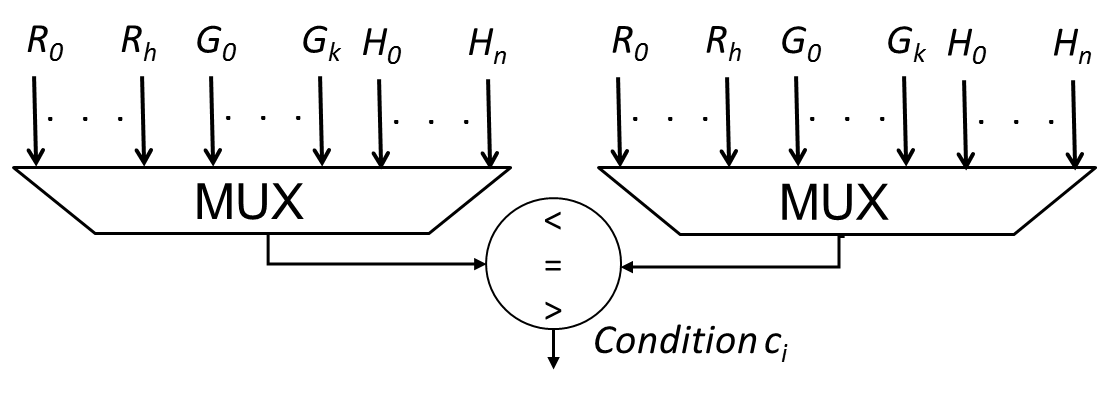}
      	  \vspace{-2em}
\caption{Condition logic block array element}
   	  \vspace{-1.8em}
\label{F:cond}
\end{figure}

\vspace{3pt} \noindent {\bf Condition Logic Block}. This block permits to configure conditions on input pairs (per-flow registries, global registries, header fields), and evaluate them so as to return as output a boolean 0/1 vector. This block, shown in figure \ref{F:cond}, comprises multiple (8 in our implementation) parallel configurable comparators, each of which takes as input two operands selected among all the flow registries $R_i$, all the global registries $G_i$ and the header fields $H_i$ coming from the packet field extractor. The selection operation is provided by two multiplexers (one for each operand). Each comparator supports five arithmetic comparison functions: $>$, $\ge$, $=$, $\le$, $<$.

\vspace{3pt} \noindent {\bf XFSM Table}. While, conceptually, this is a key ``computational'' stage in our proposed architecture (it performs a state transition step), in practice its implementation is straightforward: it just relies on an ordinary TCAM. Although a-posteriori it may seem obvious, such a simple support for a ``full'' XFSM transition step was enabled by the clear distinction between the configuration and evaluation of conditions (by the Condition Logic Block) and their usage as boolean outcomes, hence one bit per condition which can be directly used as TCAM input, along with the state label and the usual packet header fields used in OpenFlow matches. As in standard implementations, the TCAM provides as output the row associated to the matching rule with higher priority, and is followed by a companion RAM which stores the associated output. In our specific case, this consists in i) the next state label (16 bits) used to update the flow context table, ii) the action to perform on the packet (16 bits) and the ALU  instructions which shall be applied to update registries (our prototype supports up to 5 instructions of 32 bits each).

\vspace{3pt} \noindent {\bf Update Logic Block}. This is the second ``computational'' stage in our architecture. This block deploys an array of ALUs (Arithmetic and Logic Units) which support a specific set of (micro)instructions useful for traffic processing tasks, and which {\em execute} in parallel the instructions provided as output of the XFSM Table. The updated registry values are then stored in the relevant memory locations (flow registries and/or global registries). Technical details are provided in section \ref{ss:alu}.

\vspace{3pt} \noindent {\bf Miscellaneous Blocks and Microcontroller}. To complete the HW architecture, a few necessary extra blocks have been implemented. The {\em Action Block} applies the selected actions to the packet and is perfectly analogous to an OpenFlow implementation. Being just a proof-of-concept, our prototype implements only basic ``sample'' actions (drop, forward, flood). {\em Global registries} are implemented as a standard register file unit for concurrent access. The {\em Metadata block} is in charge to provide additional information associated to an arriving packet, i.e., input port and timestamp.
Finally, our prototype has been complemented with a {\em microcontroller} providing a communication interface (UART) to configure the various programmable components inside the OPP (configuration registers, TCAM and RAM memories, etc.), i.e. to deploy in the switch an externally programmed application. Each configurable quantity is memory mapped in the microcontroller address space, which can directly read/write the content of these components.  The microcontroller further implements management functions, among which slow-time-scale flow context table management (housekeeping): the microcontroller periodically scans the entries in the flow context table to detect and clean stale entries. To this purpose, two activity flag bits are stored in each flow entry and permit to label entries as \texttt{ACTIVE}, \texttt{INACTIVE} (no accesses have occurred in a configurable management cycle, e.g., order of seconds), and \texttt{DELETED}. It is worth to note that this is the only operation performed by the OFP that is not triggered by a packet.

\subsection{Update logic block}
\label{ss:alu}

Besides the support for state transition, a further key motivation behind this work was the attempt to cleanly design {\em inside the abstraction} (and concretely support in the OPP architecture) computational primitives involving arithmetic processing, as this is frequently needed in many application. The Update Logic Block is the OPP component in charge to provide such facility. It comprises a number (5 in our implementation) of small parallel ALUs (Arithmetic Logic Units) able to perform a set of elementary instructions which frequently occur in traffic control applications. The ones specifically implemented in our prototype are listed in table \ref{T:ALU} and \ref{T:specistr}. Some of these instructions are those of a typical RISC architecture, while others are specific for packet processing tasks (last row in the table). 

At each step, the specific computations that the Update Logic Block must perform are provided by the output of the XFSM transition, and are expressed in the form of a tuple of instructions (32 bit instructions in our prototype). Each instruction comprises an 8 bits $OPCODE$, followed by a variable number of operands that depend on the specific instruction. Input operands ($INi$) can be any among the available per flow registries $R_i$, the global variables $G_i$, or the header fields $H_i$ provided by the Extractor. Output operands ($OUTi$) indicate where the result of the instruction must be written (e.g. in a given per-flow register, or in a global variable). In some instructions, one or more of the operands ($IOi$) are both used as input and output. Our implementation supports 4 per-flow registries, 4 global registries and 8 header fields. Therefore, it may in principle support up to $24/\log_2(16)=6$ operands. In practice, we envision at most 4 operands (e.g., for the variance or for the ewma smoothing instructions) and thus our implementation may readily support up to 64 among registries and header fields. In the case of logic/arithmetic/shift operations, which only require at most two operands plus a third output, we have also considered the case in which one of the operands is an actual value (immediate value) which can hence use 16 bits.

\begin{table}[t]
\vspace{-.6em}
\centering
\small
\begin{tabular}{|c|c|c|}
  \hline
  Type & Instructions & definition \\
\hline
Logic & NOP & do nothing\\
   & NOT & $OUT1 \leftarrow NOT(IN1)$\\
 & XOR, AND, OR & $OUT1 \leftarrow IN1\; op\; IN2$\\
\hline
Arit. & ADD,SUB,MUL,DIV     & $OUT1 \leftarrow IN1\; op\; IN2$\\
    & ADDI,SUBI,MULI,DIVI & $OUT1 \leftarrow IN1\; op\; IMM$\\
\hline
Shift/ & LSL (Logical Shift Left) & $OUT1 \leftarrow IN1 << IMM$ \\ 
Rotate  & LSR (Logical Shift Right) & $OUT1 \leftarrow IN1 >> IMM$\\
						 & ROR (Rotate Right)  & $OUT1 \leftarrow IN1\; ror\; IMM$ \\
\hline
\end{tabular}
      	  \vspace{-1em}
\caption{ALU basic instruction set}
      	  \vspace{-1em}

\label{T:ALU}
\end{table}

The packet/flow specific instructions supported in our prototype do implement, as a dedicated HW primitives running at the system clock frequency and with a maximum latency of two clock cycles\footnote{As they involve a division, which we had to limit to 16 bits for dividend and divisor to target a 2 clock cycles latency.}, domain-specific operations which we deem useful in traffic control applications, and which would normally require multiple clock cycles if implemented using more elementary operations. Such domain specific operations include the online computation of running averages ($avg$) and variances ($var$), and the computation of exponentially decaying moving averages ($ewma$) which can serve the purpose of a moving average, but which can be incrementally computed and do not require to maintain a window of samples. 

Usage and implementation details about packet/flow specific instructions are provided in Table \ref{T:specistr}. The $avg$ operation stores the number of samples in $IO1$, and includes a new sample $IN1$ in  the running average $IO2$. Similarly, the $var$ operation stores the number of samples in $IO1$, the average of the value $IN1$ in $IO2$ and the variance in $IO3$.

The $ewma$ operation\footnote{Being $t_k$ the last sample time, and $x_{k'}$ a new sample occurring at time $t_{k'}$, for simplicity of HW implementation we {\em approximate} the exponentially weighted moving average as $m(t_{k'})=m(t_k) \alpha^{t_{k'}-t_k} + x_{k'}$, and we use $\alpha=1/2$ to compute powers as shift operations. The intermediate $decay$ quantity in the second line is used just for clarity of presentation.}
was included to permit smoothing. It stores the last timestamp ($IN1$) of a packet in the register identified by $IO1$, computes the exponentially weighted moving average of the value $IN2$ using the equation in Table \ref{T:specistr} and stores the result in $IO2$.

As a final remark, similar to the action set in standard OpenFlow, we stress that the specific instruction set provided by the Update Logic Block is independent of our proposed OPP abstraction, i.e., its extension or improvement (e.g. with further dedicated domain-specific instructions) does not affect the overall OPP design.

\begin{table}[t]
\vspace{-.6em}
\centering
\small
\begin{tabular}{|c|c|}
  \hline
 Instruction & definition \\
\hline
avg()  & $IO1 \leftarrow IO1+1$ \\
       & $IO2 \leftarrow IO2+(IN1-IO2)/(IO1+1)$\\
\hline
       & $IO1 \leftarrow IO1+1$\\
var()  & $IO2 \leftarrow IO2+(IN1-IO2)/(IO1+1)$\\
       & $IO3 \leftarrow IO3+((IN1-IO2)^2 -IO3)/(IO1+1)$\\
\hline
        & $IO1 \leftarrow IN1$ \\
ewma()  & $decay = 1<<(IN1-IO1)$ \\ 
		& $IO2 \leftarrow IO2/decay +IN2$ \\ 
\hline
\end{tabular}
      	  \vspace{-1em}
\caption{ALU packet/flow specific instructions}
      	  \vspace{-1em}
\label{T:specistr}
\end{table}

\subsection{FPGA prototype}
\label{s:proto}

The OFP HW prototype has been designed using as target development board the NetFPGA SUME \cite{sume}, an x8 Gen3 PCIe adapter card incorporating a Xilinx Virtex-7 690T FPGA \cite{V7},  four SFP+ transceivers providing four 10GbE links, three 72 Mbits QDR II SRAM and two 4GB DDR3 memories.
The FPGA is clocked at 156.25 MHz, with a 64 bits data path from the Ethernet ports, corresponding to a 10 gbps throughput per port. The aggregated bus output of the mixer is 320 bits wide and is able to provide an overall throughput of 50 Gbps. The d-left hash table implementing the flow context table is sized for 4K entries. In order to support the target throughput, the RAMs composing the d-left table are realized as dual port RAM, so as to provide a read and a write operation for each clock cycle.

The prototype implements very small TCAMs. The TCAM associated to the hash table in the flow context table has 32 entries of 128 bits, whereas the XFSM TCAM has 128 entries of 160 bits. Indeed, TCAM implementation over FPGAs is very inefficient and is currently a widely open research issue \cite{TCAM1,TCAM2,TCAM4}, especially since the priority resolution hardware limits the maximum operating frequency when the number of TCAM entries increase. It is thus more interesting to {\em understand the performance that would be achievable with an ASIC implementation}. Following the same technology assumptions of \cite{Bos13}, an OPP ASIC design would be able to work at 1GHz operating frequency. This corresponds to an aggregate throughput of 960M packets/s, that is the maximum achievable by a 64 ports 10 Gb/s switch chip. However, the most important scaling provided by the ASIC implementation is given by the number of entries that can be stored in the OPP tables. The size of the SRAM that can be instantiated on a last generation chip is up to 32 MB, corresponding to around 1 millions of entries in the d-left hash for the context flow table. The size of a TCAM can be up to 40 Mb, corresponding to 256K XFSM table entries.

The system latency, i.e. the time interval from the first table lookup to the last context update is 6 clock cycles. The FPGA prototype is able to sustain the full throughput of 40 Gbits/sec provided by the 4 switch ports. If we suppose a minimum packet size of 40 bytes (320 bits), the system is able to process 1 packet for each clock cycle, and thus up to 6 packets could be pipelined. However, the feedback loop (not present in the forward-only OpenFlow pipelines \cite{OF1.4}) raises a concern: the state update performed for a packet at the sixth clock cycle would be missed by pipelined packets. This could be an issue for packets {\em belonging to a same flow} arriving back-to-back (consecutive clock cycles); in practice, as long as the system is configured to work by aggregating $N \geq 6$ different links, the mixer's round robin policy will separate two packets coming from the same link of $N$ clock cycles, thus solving the problem. Note that the 6 clock cycles latency is fixed by the hardware blocks used in the FPGA (the TCAM and the Block RAMs) and basically does not change scaling up the number of ingress ports or moving to an ASIC.

\begin{table}[t]
\centering
\small 
\begin{tabular}{|c|c|c|}
  \hline
resource type & Reference switch & OPP switch   \\
\hline
\# Slice LUTs & 49436 (11\%) & 71712 (16\%)  \\
 \hline
\# Block RAMs & 194  (13\%)  & 393 (26\%) \\
  \hline
\end{tabular}
      	  \vspace{-0.6em}
\caption{Hardware cost of OPP compared with the reference NetFPGA SUME switch.}
      	  \vspace{-1.5em}
\label{t:synth}
\end{table}

\begin{figure}[t]
    \centering
    \includegraphics[width=0.4\textwidth]{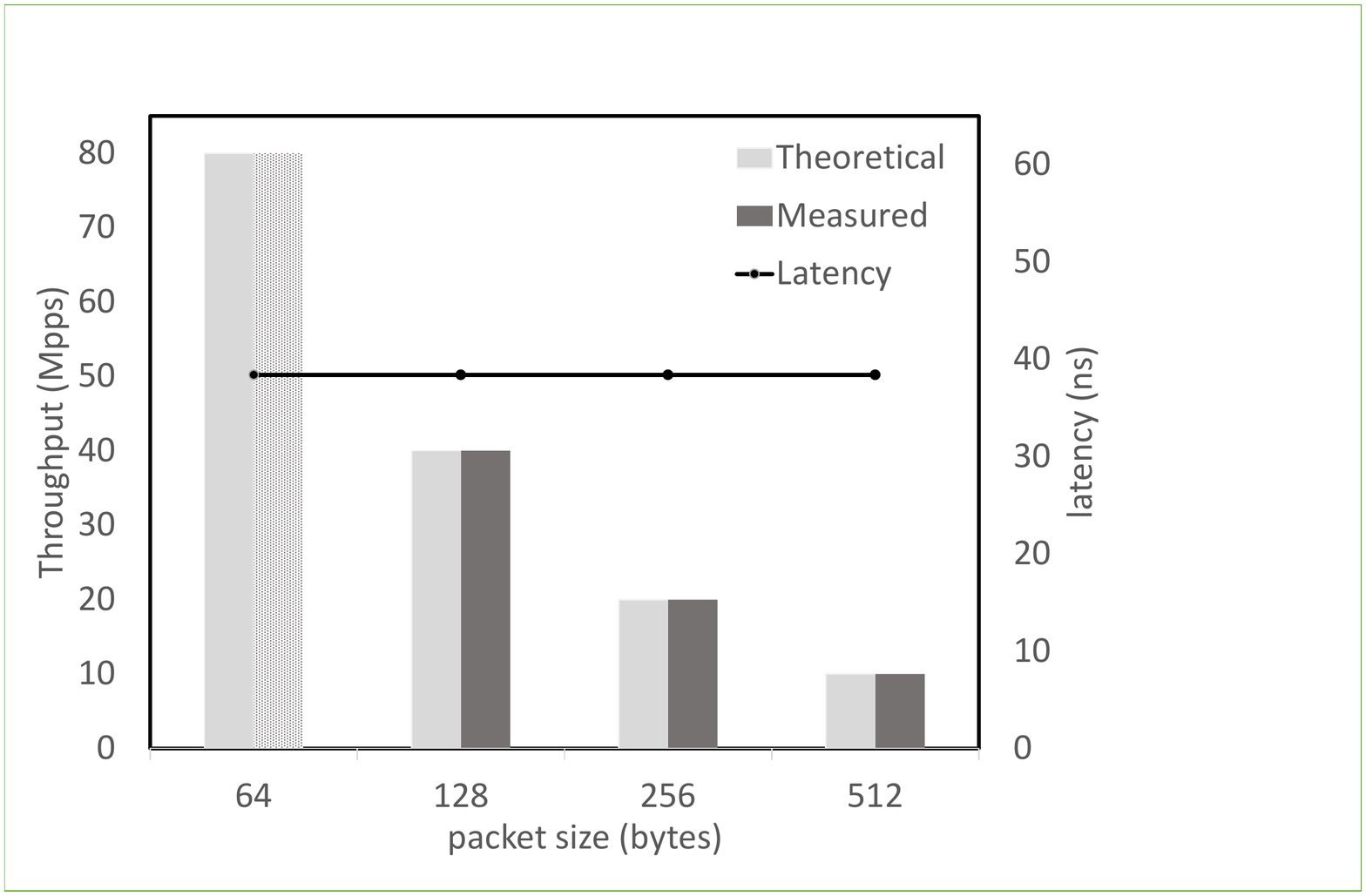}
          	  \vspace{-1em}
    \caption{Performance of the FPGA prototype}
          	  \vspace{-1.2em}
    \label{F:perfo}
 \end{figure}

The whole system has been synthesized using the standard Xilinx design flow. Table \ref{t:synth} reports the logic and memory resources (in terms of absolute numbers and fraction of available FPGA resources) used by the OPP FPGA implementation, and compare these results with those required for the NetFPGA SUME single-stage reference switch. As expected, the logic uses a small fraction of the total area (the increase with respect the reference switch is 5\% of the available FPGA logic resources), that is dominated by memory (that doubles with respect the reference switch). The synthesis results hence confirm the trend already shown by \cite{Bos13}: the HW area is dominated by memory, while adding intelligence/features in the logic require a small silicon overhead. The performance in terms of latency of an OPP stage and throughput of deployed FPGA prototype has been measured sending several synthetic traces of packets of different size. The results are presented in Fig. \ref{F:perfo}. As expected, the FPGA is able to sustain the expected throughput\footnote{Due to the limitation of our hardware measurement set-up, we were unable to actually send to the FPGA more than 24 Gbits/sec, so the data referred to 64B packet size could not be measured. The expected theoretical value is reported.}

\section{Programming examples}
\label{s:usecases}

To functionally test the ability of OPP to support stateful applications, we have developed a complete OPP virtual software environment. For both the switch and controller implementation we have extended the CPqD OpenFlow 1.3 software virtual switch \cite{ofsoftswitch13} and the widely adopted OpenFlow controller Ryu \cite{ryu}. The sotware implementation serves just for testing purpose, hence it closely mimics the described OPP hardware operation, including the relevant limitations. To configure an OPP switch via the controller, we developed an OPP-specific extension of the OpenFlow protocol. Due to space limitations (the interested reader can find configuration files in the repository) we just mention that the configuration of the XFSM in the OPP architecture is a straightforward extension of an OpenFlow configuration: it just requires to populate the XFSM table entries and to configure of conditions, functions, key extractors and initial global register values. All software components required to test the proposed applications are bundled in a mininet \cite{mininet} based virtual machine avilable at a {\em dedicated OPP repository} \cite{repository}, along with our prototype's VHDL HW code. 

\begin{figure}[t]
\centering
   \includegraphics[width=0.49\textwidth]{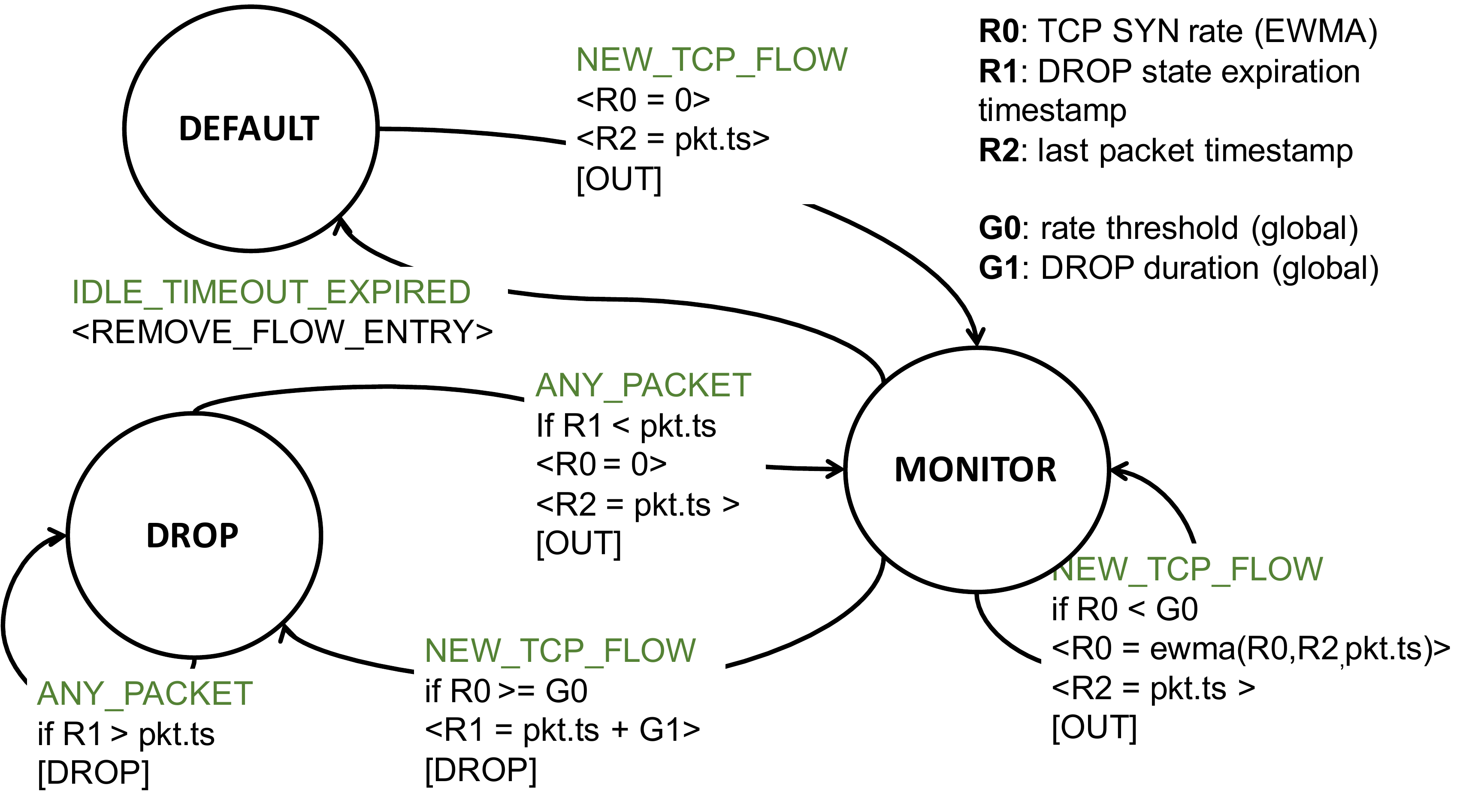}
      	  \vspace{-2.2em}
\caption{Port scan detection XFSM}
\label{F:portscan}
\end{figure}

To understand how an application can be programmed using OPP, let's walk through a simplistic example of a (quite inefficient, but at least trivial to follow) TCP port scan detection and mitigation application. Since the target is to detect IP address which behave as scanners, we use as Flow Key the IP address. Figure \ref{F:portscan} represents our desired application's behavior, expressed in the form of an XFSM ``program'', whereas figure \label{F:pscan_table} provides a corresponding tabular configuration delivered to the switch. For every IP packet, we check in the Flow Context Table whether the IP source has an associated context; if this is not the case, a \texttt{DEFAULT} state is conventionally returned. the XFSM table now checks whether the packet is a TCP SYN, and only in this case we will allocate a Flow Context Table entry for the considered IP source, and we will set it in \texttt{MONITOR} state. In this state, we measure the rate of new TCP SYN arrivals toward hosts behind the switch port 1. Such rate (computed with the EWMA update function) is stored and updated in the flow registers $R_0$. 

While in \texttt{MONITOR} state, the value of $R_0$ is verified for each new TCP flow. If a given threshold (say 20 SYN/s, a value stored in the global register $G_0$) is exceeded, the state associated to this flow is set to \texttt{DROP} and all packets from this IP addresses are discarded. Suppose now that the programmer wants to block the scanner for 5 seconds. Lacking explicit timers (a non trivial HW extension), such mechanism is realised by the following procedure:  (i) when the flow state transits from \texttt{MONITOR} to \texttt{DROP} the register $R_1$ is set to the packet time stamp value plus 5 sec. (a value stored in the global register $G_1$); (ii) in \texttt{MONITOR} state the $R_1$ value is checked for every received packet; (iii) If $R_1 <= pkt.ts$ the flow state is reverted to \texttt{MONITOR}. Moreover, the application needs to store the last packet timestamp in the flow register $R_2$. This time stamp will be used by the EWMA update function. 

\begin{figure}[t]
\centering
   \includegraphics[width=0.49\textwidth]{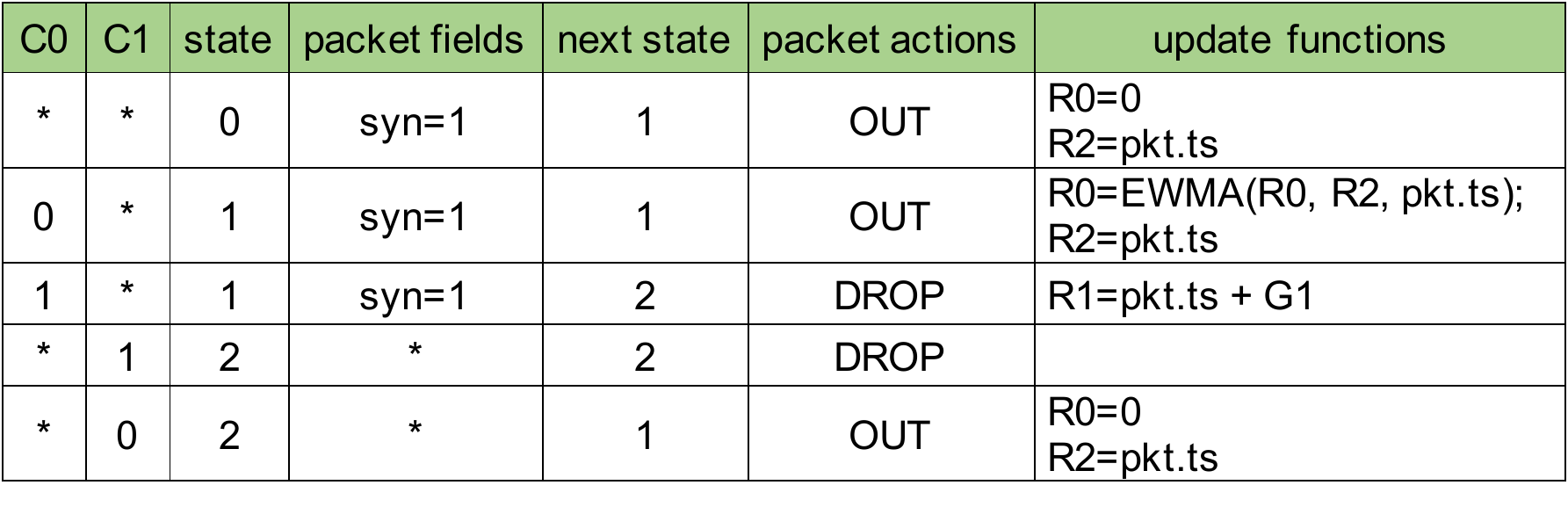}
         	  \vspace{-2em}
\caption{Port scan detection XFSM table}
      	  \vspace{-1.5em}
\label{F:pscan_table}
\end{figure}

To implement this application, the OPP switch is configured by the controller using the OPP protocol that performs the following elementary operations: i) the \texttt{DEFAULT},  \texttt{MONITOR} and  \texttt{DROP} states are encoded with 0, 1 and 2 respectively; ii) the lookup and update scopes are set to \texttt{ip.src}; iii) the global registers are defined as follows: $G_0 = 20$ and $G_1 = 5$[s]; iv) the condition table is configured to include two conditions: $C_0: R_0 \ge G_0$ and $C_1: R_1 > pkt.ts$ (packet arrival time); v) the XFSM table is configured as in fig. \ref{F:pscan_table}.

\subsection{Decision tree based traffic classification}
Machine Learning (ML) tools are widely adopted by the networking  community for detecting anomalies and classifying traffic patterns \cite{nguyen2008survey}. We have tested the feasibility of using OPP to support this kind of traffic monitoring schemes by implementing a decision tree supervised classifier based on the C4.5 algorithm \cite{quinlan2014c4} that has been exploited by different work on ML based network traffic classification \cite{williams2006preliminary,pan2003hybrid,ma2008study,zhang2010method,li2007machine,alshammari2009machine}.

Any ML based classification mechanisms has two phases: a {\em training phase} and a {\em test phase}. The training phase is off line and used to create the classification model by feeding the algorithm with labeled data that associate a measured traffic feature vector to one of {\em n} decision classes. In the case of decision tree based ML algorithms, the output of such phase is the {\em binary classification tree}. The training phase must obviously be performed outside the switch. For our use case implementation the decision rules have been created using the Orange data mining framework \cite{JMLR:demsar13a}, and a feature set proposed in  \cite{li2007accurate}. We considered a simple scenario where it is necessary to discriminate between WEB and P2P (control) traffic. The selected features for each flow are: packet size average/variance and total number of received bytes. These features are mapped directly to the per flow memory registers R1, R2, R3. Moreover the application XFSM requires two additional registers: R0 (packet counter) and R4 (measurement window expiration time). The input feature vectors are evaluated over a time window of 10 seconds.

\begin{figure}[t]
   \centering
   \includegraphics[width=.45\textwidth]{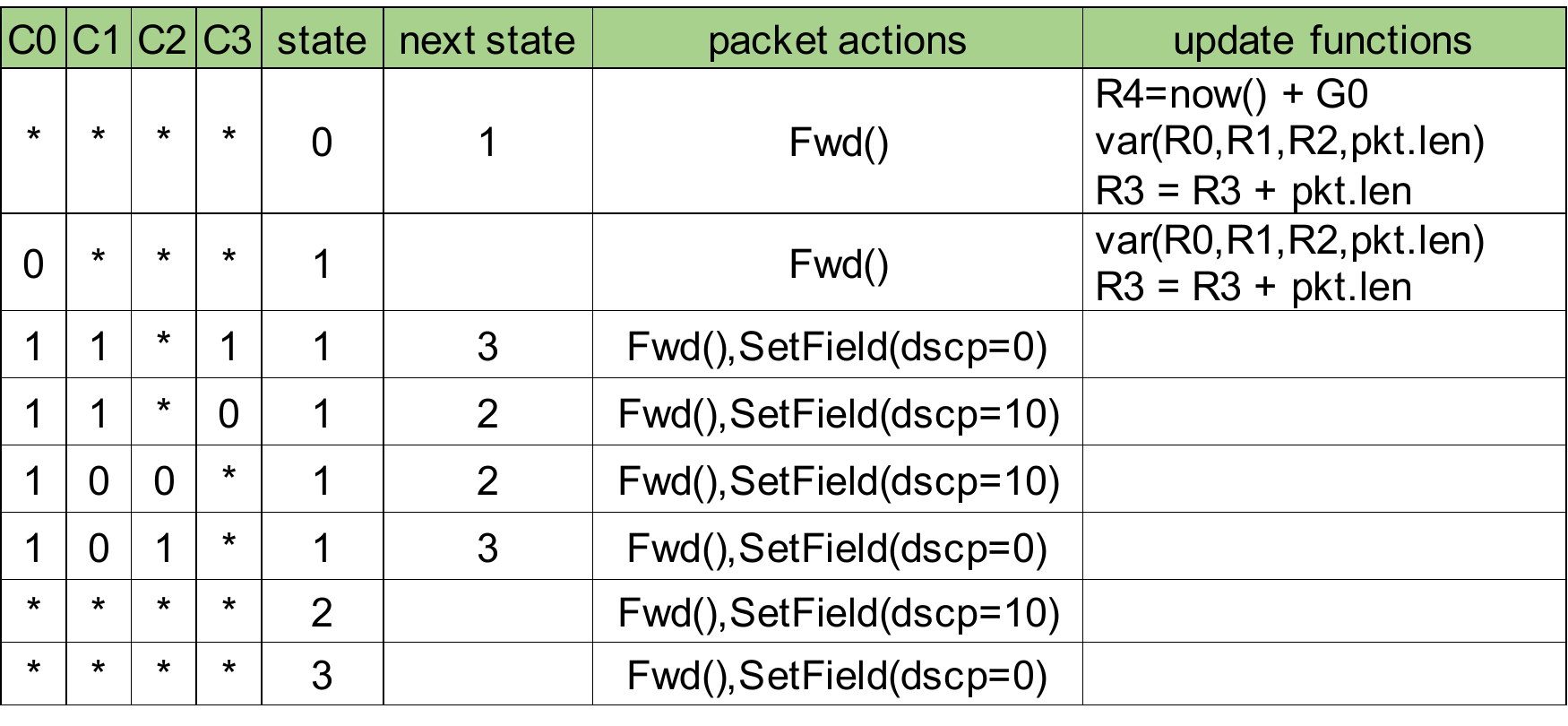}            	  \vspace{-1em}

   \caption{XFSM table for the traffic classifier}
            	  \vspace{-1.5em}
   \label{fig:tree}
\end{figure}

The test phase, which is performed online, consists of two operations: (1) for each flow the feature set described above is computed; (2) after 10 seconds a decision is made according to the decision tree. This testing mechanism is implemented in OPP according to the XFSM shown in Figure \ref{fig:tree}. The XFSM flows states are encoded as follows: State 0 $\rightarrow$ default; State 1 $\rightarrow$ measurement and decision; State 2 $\rightarrow$ WEB traffic (DSCP class AF11); State 3 $\rightarrow$ P2P traffic (DSCP class best effort). The condition set is:  $C0: (now > R4)$;  $C1: (R2 > G2)$, $C2: R3 > G3$, $C3: R1 <= G1$. According to our simplified training phase, the (ceiled) thresholds  values are: $G1 = 306$; $G2= 1575$ and $G3 = 203$. 

\subsection{Traffic policing with token buckets}
In this second use case we have implemented in OPP a single rate token bucket with burst size $B$ and token rate $1/Q$, where $Q$ is the token inter arrival time. Since in the OPP architecture the update functions are performed after the condition verification, we cannot update the number of tokens in the bucket based on packet arrival time before evaluating the condition (token availability) for packet forwarding. For this reason we have implemented an alternative and  equivalent algorithm based on a time window. For each flow a time window W ($T_{min} - T_{max}$) of length $BQ$ is maintained to represent the availability times of the tokens in the bucket. At each packet arrival, if arrival time $T0$ is within W (Case 1), at least one token is available and the bucket is not full, so we shift W by Q to the right and forward packet. If the arrival time is after $T_{max}$ (Case 2), the bucket is full, so packet is forwarded and W is moved to the right to reflect that $B-1$ tokens are now available ($T_{min}=T0-(B-1)Q$ and $T_{max}=T0+Q$). Finally, if the packet is received before $T_{min}$ (Case 3), no token is available, therefore W is left unchanged and the packet is dropped.

In the OPP implementation, upon receipt of the first flow packet, we make a state transition in which we initialize the two registers: $Tmin=T0-(B-1)*Q$ and $Tmax=T0+Q$ (initialization with full bucket). 

At each subsequent packet arrival we verify two conditions: $C0: Tnow>=Tmin$; $C1: Tnow<=Tmax$.

The three cases defined by the algorithm can be easily identified with these two conditions: case 1) C0 == True AND C1 == True; case 2) C0 == True AND C1 == False; case 3) C0 == False.

\begin{figure}[t]
	\centering
	\includegraphics[width=.45\textwidth]{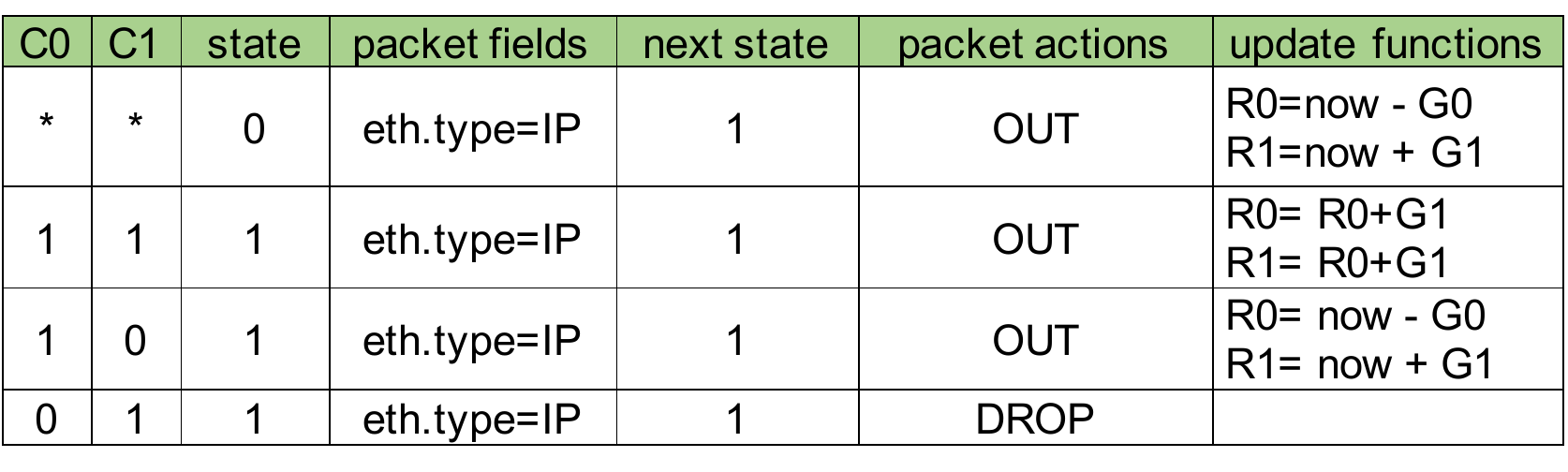} 
	         	  \vspace{-1.2em}
	\caption{Token bucket XFSM. The flow registers $R0$, $R1$ are used to store respectively Tmin, Tmax. The global registers $G0, G1$ are used to store $B*Q$ and $Q$. The extractors are $ip.src$}
	         	  \vspace{-1.5em}
	\label{fig:tok_xfsm}
\end{figure}

The XFSM is shown in Figure \ref{fig:tok_xfsm}. 

\section{Discussion and extensions}
\label{s:disc}

\noindent \textbf{Hard-coded vs. programmable features}. \\
The reader might have noted that, in the proposal of the XFSM abstraction and in the technical design of OPP, we have attempted to {\em not rely} on {\em any} stateful feature today available in OpenFlow switches. This might seem odd, as, for a very basic example, per-flow statistics are assumed since \cite{OF08} to be collected and implemented by the switch hardware itself, and thus would be readily available in virtually any baseline reference OpenFlow switch architecture. However, at least for the purpose of this work, we have voluntary avoided to {\em expose}, via our abstraction, features (such as OpenFlow-type statistics) that could at least in principle be programmed from the outside, by {\em using} the abstraction.  Thus, in the attempt to keep our proposed abstraction as clean as possible, we avoided focusing on possible optimizations, such as in-switch efficient implementation of frequently used features (such as per flow count/byte statistics, soft states and timers, etc.) which one would arguably expect from an advanced implementation, so as to concentrate programming efforts on one own application's logic. 

\vspace{3pt} \noindent \textbf{Proposed abstraction: at which ``level''?}. \\
Our abstraction resides at a very low level, as in practice it specifies a machine language which directly configures the hardware. In this, albeit technically very different, we are quite similar in spirit to the original OpenFlow abstraction, envisioned as an abstraction of the switch's Flow table component. Independence of the underlying platform is therefore {\em not} accomplished via a compiler, such as in the case of emerging data plane programming languages such as P4 \cite{Jos15}, but it is accomplished by {\em decoupling} the actual identification and (XFSM-based) combination of the hardware primitives from how they are implemented inside the switch (the proposed OPP hardware design being a possible implementation, but not nearly the unique). As long as two platforms expose a same set of (header) matching facilities, forwarding actions, enabling functions for evaluating conditions, and a same instruction set for the update functions, and as long as they do support the XFSM state transition execution logic, the application's description in terms of XFSM can be ported across platforms. Loosely speaking, our abstraction is more closer to a ``bytecode'' or to an assembly language rather than to a practical programming language. While we believe this is a strength of our proposal, we also clearly recognize the importance of offering to the applications' developers an higher level and more user-friendly programming language (P4 being an obvious candidate, see next discussion) and relevant compilers which transform an higher level description into the proposed OPP XFSM-based machine language. 

\vspace{3pt} \noindent \textbf{Use of P4 for the description of OPP stages.} \\
Even if the details of the reference P4 hardware architecture is not publicly available, the language has the goal of being expressive enough to describe any configurable architecture for packet forwarding. Therefore, even if out of the scopes of our paper, and for just the purpose to stimulate possible discussion (i.e. we don't claim this paper to provide any specific contribution in this direction), we nevertheless made our own preliminary attempt to understand how the proposed OPP architecture could be described using P4, i.e. if OPP could be used as a further platform's target for a P4 program. To this purpose, we have defined (and uploaded over the anonymized OPP's repository) a reusable OPP.p4 library. From such experience, we gathered a twofold impression. On one side, the current version of the language already permits to describe/support key OPP functionalities. Moreover, the ability to store data in persistent registries permits to properly describe, using P4, the ``computational loop'' characterizing OPP. 
On the other side, we suffered from the lack, among the P4 constructs, of an explicit state/context table and a relevant clean way to store and access per-flow data. In our OPP.p4 library, a table functionally equivalent to our Context table was actually constructed by combining arrays of registers with hash keys generators which are provided as P4 language primitives. However, besides the obvious stretch (P4 registers are generic, and not specifically meant to be deployed on a per-flow basis), this construction also suffers from hash collisions, a non trivial problem if constrained to be addressed {\em while} the packet is flying through the pipeline. The availability of a tailored context/state table structure in P4 would greatly simplify the support of an OPP target platform.

\vspace{3pt} \noindent \textbf{Structural limitations and possible extensions} \\
While (we believe) very promising, our proposed approach is not free of {\em structural} concerns. If, on one side, limitations in the set of supported enabling functions and ALU functions for registry updates may be easily addressed with suitable extensions, and integration of more flexible packet header parsing (following \cite{Bos14}) is not expected to bring significant changes in the architecture, there are at least three pragmatic compromises which we took in the design, and which suggest future research directions. The first, and major, one resides in the fact that state transitions are ``clocked'' by packet arrivals: one and only one state transition associated to a flow can be triggered only if a packet of {\em that} flow arrives; asynchronous events, such as timers' expiration, are not supported. So far we have partially addressed this limitation with, on one side, the decoupling between lookup and update functions (the cross-flow state handling feature), and on the other side with programming tricks such as the handling of time performed while implementing the token bucket example. But further flexibility in this direction is a priority in our future research work.  A second shortcoming is the deployment of ALU processing only in the Update Logic Block. This decision was done in favour of a cleaner abstraction and a simpler implementation. However (programmable) arithmetic and logic operations would be beneficial also {\em while} evaluate conditions ($e.g.$, $A-B > C$) which, in the most general case, may require to be postponed to the next packet (the update function can store $A-B$ in a registry, and the next condition can use such registry). A third, minor, shortcoming relates to the fact that all updates occur in parallel. This prevents the programmer to pipeline operations, i.e. use (in the same transition step) the output of an instruction as input to a next one. While this issue is easily addressed by deploying multiple Update Logic Blocks in series, this would increase the latency of the OPP loop.


\section{Related work}
\label{s:related}

This work focuses on data plane programming architectures and abstractions, a relatively recent trend. In such field, so far the mostly influential work is arguably P4 \cite{Bos14} a programming language specifically focusing on data path packet processing. In turns, such initial work has stimulated the creation of a consortium (p4.org) which has so far produced a release 1.0.2 of the language specification \cite{P4spec}. Our OPP work is at a different (lower) level than P4: it describes an hardware programming interface and a relevant architecture which could be in future adapted to be used as compilation target \cite{Jos15} for P4. Furthermore, our work deals with stateful processing {\em across different packets of a flow}, and as such it appears perfectly complementary to the original P4 proposal \cite{Bos14} which initially focused mainly on programming flexibility of the packet pipeline (P4 registers have been introduced in \cite{P4spec}). 

Concerning stateful processing, the work closer to ours is OpenState \cite{ccr14}, and in part FAST \cite{Mos14}. With respect to OpenState, OPP makes a very significant step forward, as we support full eXtended Finite State Machines (XFSM, as defined in \cite{Che93}) opposed to the much simpler OpenState's Mealy Machines, and hence we significantly broaden the variety of applications that can be programmed on the switch. Such step requires additional new specialized hardware blocks with respect to OpenState which instead requires only marginal extensions to an OpenFlow hardware design \cite{Pon15}. 

Finally, OPP shares some technical similarities with \cite{Bos13} and with the Intel Flexpipe architecture \cite{flexpipe}, especially for what concerns the handling of ALUs in the packet processing pipeline. However, both OPP focus (on stateful processing) and architecture design remain extremely different from both \cite{Bos13} and \cite{flexpipe}. Indeed, an advised extension consists in extending OPP to handle multiple pipelined stages and hence exploit the TCAM reconfigurability concepts introduced in \cite{Bos13}. 


\section{Conclusions}
\label{s:conclusions}
OPP is an attempt to find a pragmatic and viable balance between platform-independent HW configurability and data plane (packet-level) programming flexibility. While permitting programmers to deploy more sophisticated stateful forwarding tasks with respect to the basic OpenFlow's static match/action abstraction, we believe that an asset of our configuration interface resides in the fact that it does not significantly depart from OpenFlow-type configurations - our extended finite state machine model is indeed  conveyed to the switch in the usual form of a TCAM's Flow Table. We thus hope that our work might stimulate further debate in the research community on how to incrementally deploy programmable traffic processing inside the network nodes, e.g. via gradual OpenFlow extensions. 

\newpage
\bibliographystyle{IEEEtran}

\bibliography{biblio}
\end{document}